\documentclass[a4paper,11pt]{article}
\usepackage{pos}
\usepackage{newtxtext}
\usepackage{newtxmath}

\title{On the PB Sudakov: NNLL coefficient, CS kernel and intrinsic-kt. }
 \ShortTitle{On the PB Sudakov}

\author*[]{Aleksandra Lelek}

\affiliation[]{University of Antwerp,\\
    Belgium}

\emailAdd{aleksandra.lelek@uantwerpen.be}

\abstract{The Transverse Momentum Dependent (TMD) Parton Branching (PB) method incorporates elements of TMD physics into a Monte Carlo (MC) framework to produce high-energy QCD predictions for collider processes. It derives TMDs from the PB evolution equation - solvable with MC techniques - fits them to
 data, and then enables their use in MC event generators for QCD predictions. In this article, we describe the 
relation of the PB Sudakov form factor to the 
Sudakov factor in the CSS formalism. We discuss both perturbative and non-perturbative components. We present recent developments to include NNLL coefficient in the PB Sudakov. We discuss the CS kernel extractions for different evolution scenarios. We remark the recent studies on intrinsic-kt vs center-of-mass (in)dependence in different approaches and their relation to the non-perturbative Sudakov. }

\FullConference{QCD at the Extremes (QCDEX2025)\\
1–5 Sept 2025\\
Online\\}

\begin{document}
\maketitle

Factorization and 
evolution of Transverse Momentum Dependent 
(TMD) parton distributions  
\cite{Angeles-Martinez:2015sea,Abdulov:2021ivr}  
enable us to incorporate soft-gluon resummations 
in high energy QCD predictions for hadron collider 
observables, increasing the accuracy and 
reliability of the predictions. 
In this article, the Sudakov form factor of the TMD Parton Branching (PB) method \cite{Hautmann:2017xtx,Hautmann:2017fcj} will be discussed and its relation to the  Sudakov of  Collins-Soper-Sterman (CSS) \cite{Collins:2011zzd} will be described, summarizing  \cite{Martinez:2024mou}.


\paragraph{The soft gluon resolution scale and AO}
The standard PDFs are obtained from DGLAP evolution, with the soft gluon resolution scale  $z_M\approx 1$. 
However, when one considers $q_0$  being the minimum transverse momentum of the emitted parton with which emitted parton can be resolved, then from the angular ordering (AO) of Catani-Marchesini-Webber, one obtains a condition on $z_M$, which becomes {\it{dynamical}}:
$
  z_{\text{dyn}}(\mu^\prime) = 1 - q_0/\mu^{\prime}
$. 
The baseline collinear MC generators apply dynamic $z_M$ in their PSs algorithms, still using PDFs with fixed $z_M \approx 1$ as an input guiding their evolution. The discrepancy between the forward evolution of the PDFs and backward evolution in the shower can affect precision of the obtained predictions \cite{Nagy:2020gjv,Dooling:2012uw}. \\
In the PB approach, instead of truncating the phase space at $z_{\text{dyn}}$, we use it as an intermediate scale to 
decompose the Sudakov form factor in two regions \cite{Martinez:2024mou}: (i) perturbative (P) for $z < z_{\text{dyn}}$, which corresponds to $|q_{\bot}| > q_0$ and (ii) non-perturbative (NP) $z_{\text{dyn}} < z < z_M$ ($z_M=1-\epsilon$ with $0 \simeq \epsilon \ll 1$), for which $|q_{\bot}|< {q_0}$:
\begin{eqnarray}
\label{eq:divided_sud}
&&\Delta_a^{} ( \mu^2 , \mu^2_0 ) =  
\exp \left(  -   
\int^{\mu^2}_{\mu^2_0} 
{{d \mu^{\prime 2} } 
\over \mu^{\prime 2} } \left[
 \int_0^{z_{\text{dyn}}(\mu')} dz 
  \frac{k_a(\alpha_s^{\rm{}})}{1-z} 
- d_a(\alpha_s^{\rm{}}) \right]\right)\nonumber \\
 & \times&  \exp \left(  -   
\int^{\mu^2}_{\mu^2_0} 
{{d \mu^{\prime 2} } 
\over \mu^{\prime 2} } 
 \int_{z_{\text{dyn}}(\mu')}^{z_M} dz 
  \frac{k_a(\alpha_s^{\rm{}})}{1-z} 
\right) =\Delta_a^{ (\text{P})}\left(\mu^2,\mu_0^2,q_0\right)  \cdot \Delta_a^{ (\text{NP})}\left(\mu^2,\mu_0^2,\epsilon,q_0^2\right)\;,
\end{eqnarray}
see e.g. \cite{Hautmann:2017fcj,Martinez:2024mou} for the detailed explanation of notation.
This modelling allows to discuss the resummation accuracy of PB and its non-perturbative component from the evolution.

\paragraph{Perturbative resummation}
\label{PertRes}
After some algebraic manipulations \cite{Martinez:2024mou}, the perturbative PB Sudakov  can be written as
$
    \Delta_a^{(\text{P})}(\mu^2,q_0^2) = \exp \left( - \int_{q_0^2}^{\mu^2} \frac{dq_{\bot}^2}{q_{\bot}^2}\left[ \frac{1}{2}k_a(\alpha_s^{\rm{}}) \ln \left(\frac{\mu^2}{q_{\bot}^2} \right) - d_a(\alpha_s^{\rm{}}) \right] \right)
$ which coincides visually with the perturbative part of the Sudakov of CSS (with appropriate scale choices) in so-called CSS1 notation. \\
The following pattern is arising from the order by order  comparison of the PB/DGLAP coefficients $k_a$ and $d_a$ with the perturbative coefficients $A_a$ and $B_a$ of the CSS Sudakov. 
 Leading order (LO) splitting functions give the leading logarithmic (LL) accuracy via
$k_a^{(0)}=A_a^{(1)}$  and  the single logarithmic term of the next-to-leading logarithmic (NLL) accuracy with $d_a^{(0)}=-\frac{1}{2}B_a^{1}$ \footnote{In PB, two separately evolved TMDs are matched with a matrix element to get the final cross section whereas in the CSS notation, the exponents from the evolution of each TMD are combined in one common Sudakov; this explains the difference of $\frac{1}{2}$. };
the next-to-leading order (NLO) splitting functions give the double logarithmic  term at NLL $k_a^{(1)}=A_a^{(2)}$, and the single-log part of NNLL by the $d_a^{(1)}$ coefficient, here however the  scheme dependence arising from the renormalization group equation has to be taken into account. The  difference between the DGLAP $d_a^{(1)}$ and the standard CSS $B_a^{(2)}$
is 
$
 B_q^{(2) \rm{DY}} -(-2)\cdot d_q^{(1)} = 16 C_F \pi \beta_0\left(\zeta_2-1\right)   
$
and
$
 B_g^{(2) \rm{H}} -(-2)\cdot d_g^{(1)} = 16 C_A \pi \beta_0\left(\zeta_2+\frac{11}{24}\right)   
$
. 


With the next-to-next-to-leading order (NNLO) DGLAP splitting functions, the pattern is broken: $k_a^{(2)}$ does not correspond to the CSS double logarithmic coefficient at NNLL, $A_a^{(3)}$, what is referred to as the collinear anomaly \cite{Becher:2010tm}. The difference    $ A_a^{(3)} - k_a^{(2)} = C_a\pi\beta_0\left[ C_A\left(\frac{808}{27} - 28\zeta_3\right) - \frac{112}{27}N_f \right]  $ is related to CS kernel via $A_a^{} - k_a^{} = -\frac{\textrm{d}\widetilde{K}_a(b_{\star}, \mu_{b_{\star}})}{\textrm{d}\ln b^2_{\star}} $. 

\begin{figure}
\begin{center} 
\includegraphics[width=4cm]{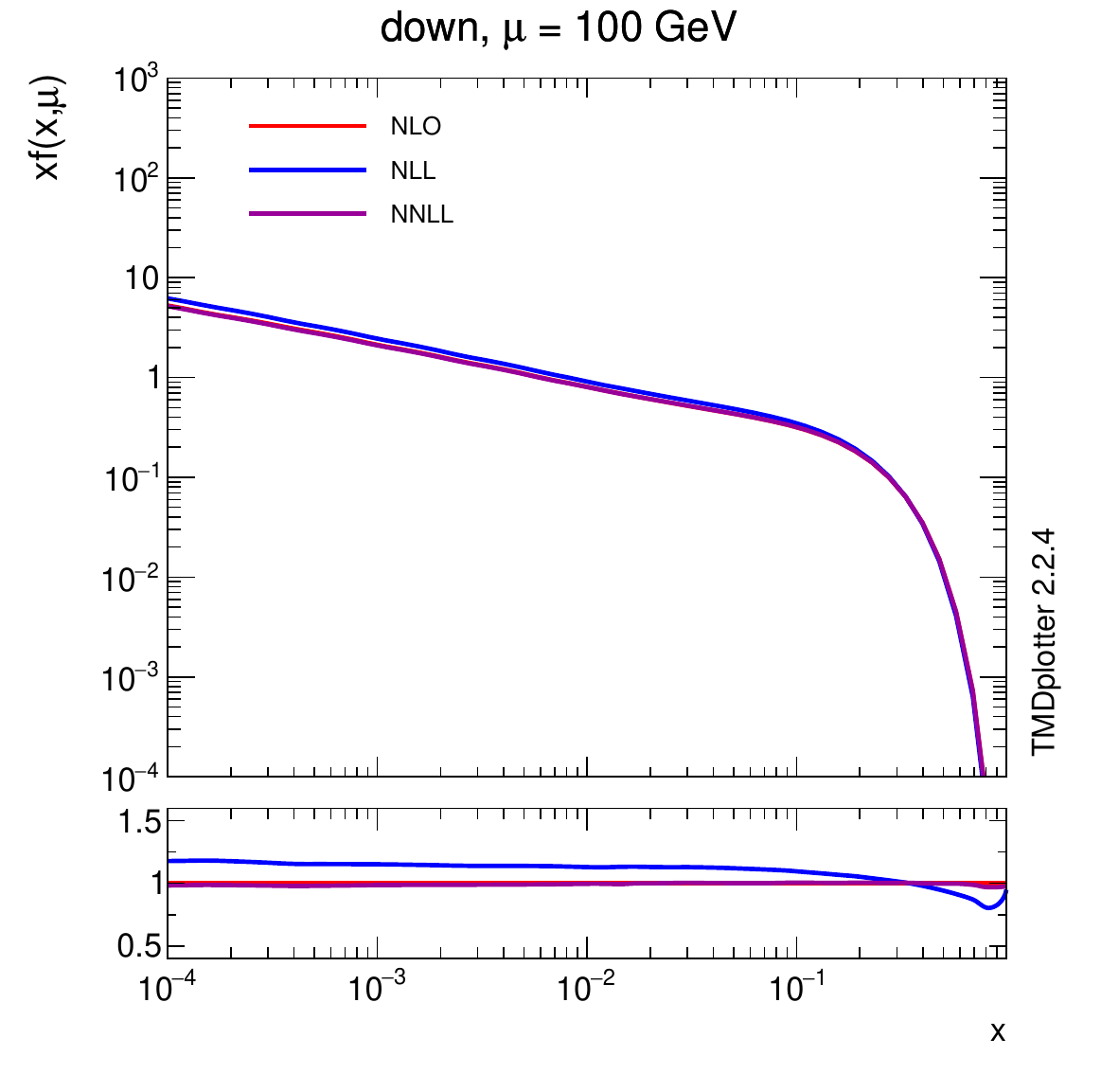}
\includegraphics[width=4cm]{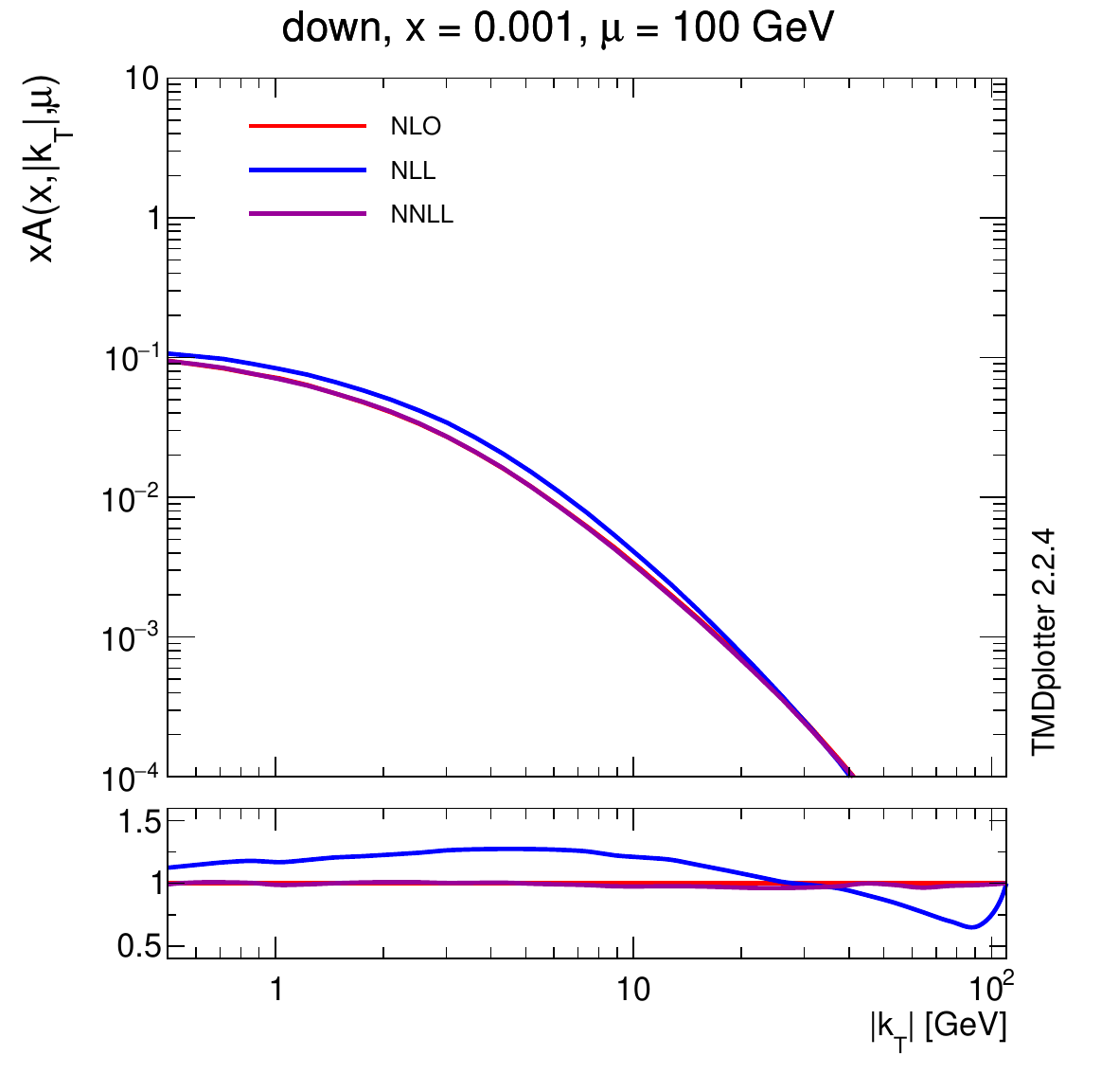}
\includegraphics[width=4.2cm]{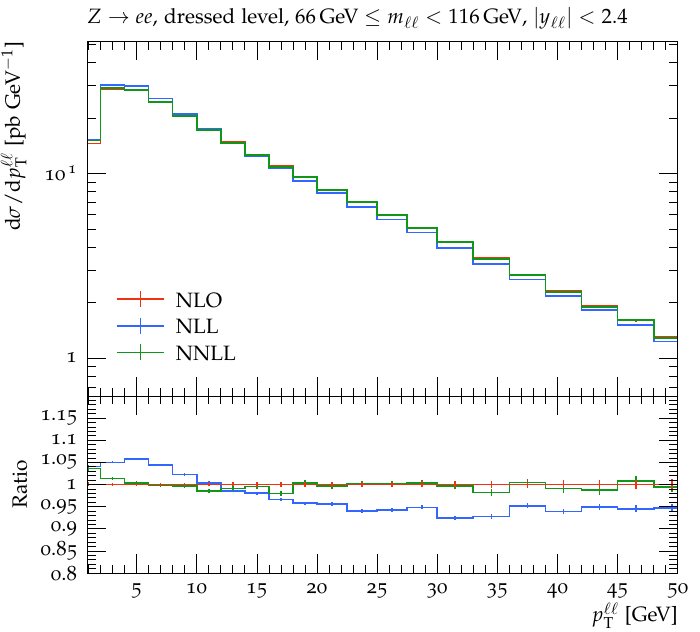}
\caption{The PB iTMDs (left) and TMDs (middle) obtained with NLO, NLL and NNLL evolution for down quark at $100\;\textrm{GeV}$ (and $x=0.001$ for the TMD case). 
The prediction for $Z$ boson $p_{\bot}$ at $8\;\textrm{TeV}$ (right) obtained with NLO ME matched to the TMDs shown in the middle plot.}
\label{fig:iTMDswithNLLNNLL}
\end{center}
\end{figure}

As proposed in \cite{Catani:2019rvy,Banfi:2018mcq},  by replacing the strong coupling by the physical (effective) soft-gluon coupling 
$
    \alpha_s^{\text{phys}}=\alpha_s \left( 1 + \sum_{n=1}^\infty \mathcal{K}^{(n)} \left( \frac{\alpha_s}{2\pi}\right)^n  \right)\
$ the NLL and NNLL accuracy of the Sudakov form factor can be reached with a proper combination of the splitting functions. 
This method was implemented in the PB evolution equation \cite{Martinez:2024mou}, and NLL and NNLL solutions were obtained. 
The results are shown in  left and middle panel of 
 Fig. ~\ref{fig:iTMDswithNLLNNLL} for the TMDs and iTMDs obtained with the 
following evolution settings:
1.) NLO:  NLO splitting functions + 2-loop $\alpha_s$; 2.) NLL: 
LO splitting functions + 2-loop $\alpha_s^{\text{NLL}}=\alpha_s\left(1 + \mathcal{K}^{(1)} \left(\frac{\alpha_s}{2\pi}\right) \right)$ 3.) NNLL:
NLO splitting functions + 2-loop 
$
    \alpha_s^{\text{NNLL}}=\alpha_s \left( 1  + \mathcal{K}^{(2)}\left( \frac{\alpha_s}{2\pi}\right)^2 \right)$\footnote{At NNLL, the piece with soft-gluon coupling $\mathcal{K}^{(1)}$ is subtracted  since it is proportional to $k_a^{(1)}$ and included via NLO splitting functions.}. For all curves, $z_M=1-10^{-5}$ and $\alpha_s(q_{\bot}^2)$  is used, $q_0=1.0\;\rm{GeV}$, for $|q_{\bot}|<1.0\;\rm{GeV}$ $\alpha_s$ is frozen to the value $\alpha_s(q_0)$.  The physical coupling has been implemented both in the Sudakov form factor as well as in the real emission probabilities to ensure momentum conservation \cite{Hautmann:2022xuc}. 
The difference between NLO and NLL  is significant both for TMDs and iTMDs, whereas the difference between the NLO and NNLL predictions is of the order of few $\%$. 

To investigate the impact of NLL and NNLL evolution on $Z$ boson $p_{\bot}$ spectrum at LHC, \textsc{CASCADE3} \cite{CASCADE:2021bxe} generator was used to match \cite{BermudezMartinez:2019anj} the TMDs from Fig.\ref{fig:iTMDswithNLLNNLL} to the  NLO matrix element generated with iTMD PB-NLO-2018-Set2 \cite{BermudezMartinez:2018fsv} with MCatNLO. 
From the right panel of Fig. ~\ref{fig:iTMDswithNLLNNLL} one can see that
there is a large difference between the NLL and NLO results and the effect of going from NLO to NNLL is of the order of few $\%$, visible in the lowest pt region.

\paragraph{Non-perturbative Sudakov}
In the non-perturbative region, the argument of $\alpha_s^{\rm{}}$  is set to $\alpha_s^{\rm{}}(q_0)$. 
With that, the non-perturbative PB Sudakov is 
\begin{eqnarray}
\label{eq:non-pert_sud}
 \ln  \Delta^{(\text{NP})}_a (\mu^2, \mu_0^2,\epsilon, q_0^2) &=& 
 - \frac{k_a(\alpha_s^{\rm{}})}{2} \ln \left(\frac{\mu^2}{\mu_0^2}\right) \ln \left(\frac{q_0^2}{\epsilon^2 \mu_0 \mu}\right) \;\;.
\end{eqnarray}
The logarithm of $\mu^2/\mu_0^2$ corresponds to the non-perturbative part of the Collins-Soper (CS) kernel. 

\begin{figure}
\begin{center} 
\includegraphics[width=5.cm]{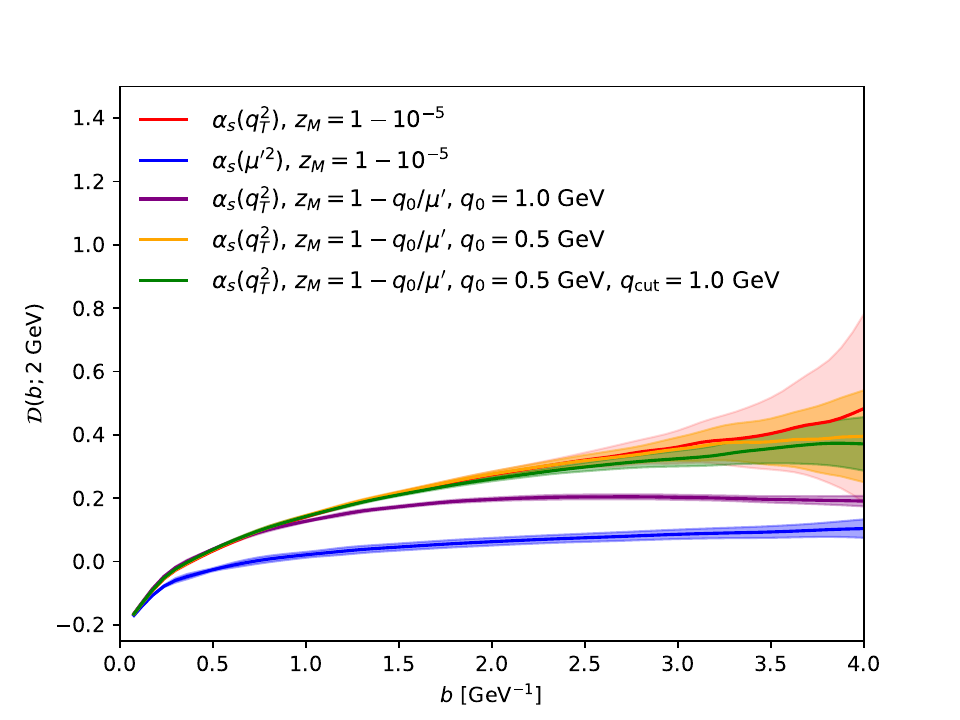}
\includegraphics[width=4.8cm]{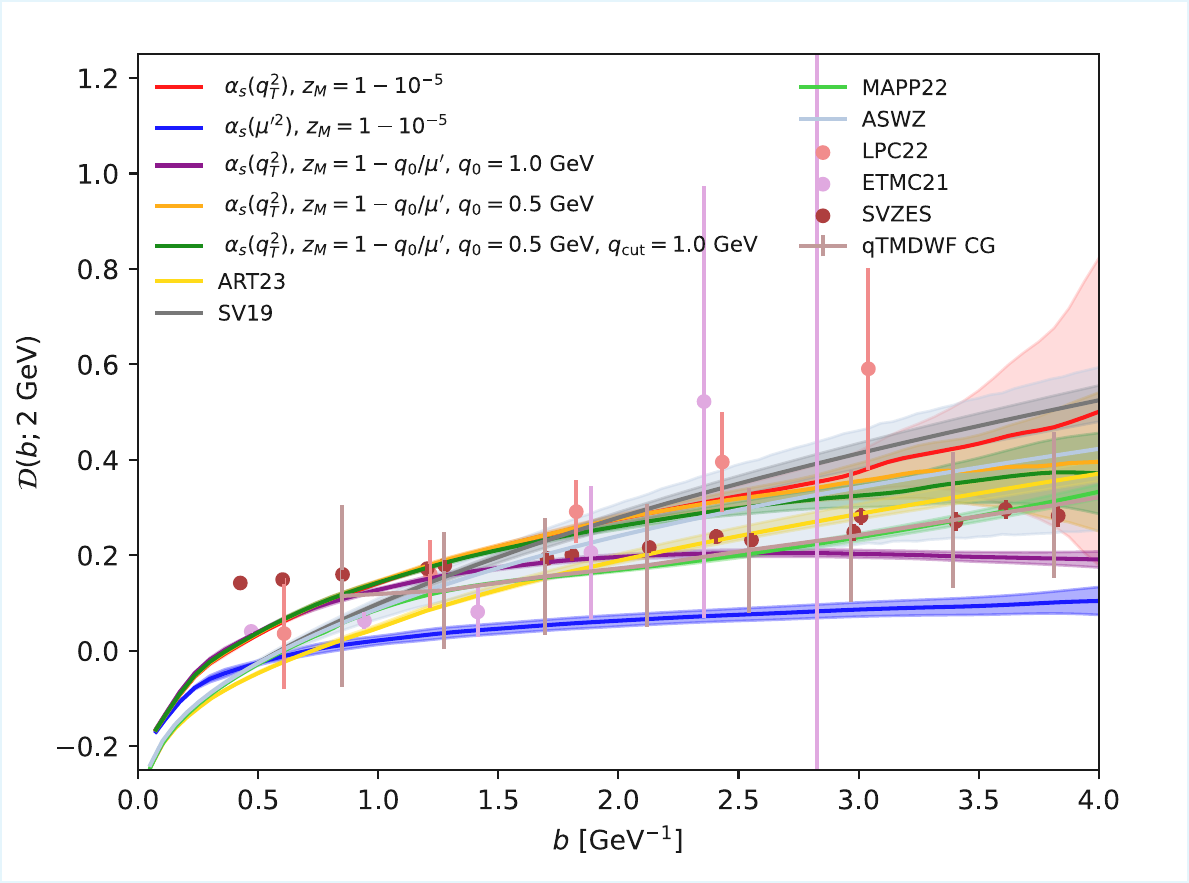}
\caption{CS kernels extracted for different PB evolution scenarios, compared to literature examples.} 
\label{fig:CSkernel}
\end{center}
\end{figure}

We use the method of \cite{BermudezMartinez:2022ctj} to extract CS kernels from 5 phenomenological models differing in the amount of (soft) radiation, modelled in terms of $\alpha_s$ and $z_M$, as explained in the legend of Fig.~\ref{fig:CSkernel}.  The results show sensitivity to the treatment of radiation: the model with the biggest amount of radiation (red)  is linear at large $b$. When the soft region is limited by neglecting the non-perturbative Sudakov (i.e. dynamic $z_M$), the kernel  becomes flatter at large $b$, the model with large $q_0=1$ GeV shows flat behaviour (purple). 
The model with $\alpha_s(\mu')$ has a contribution from very soft radiation ($z_M=1-10^{-5}$) but different scale of the coupling leads to very different kernel shape.\\
The PB predictions are compared to a selection of results from the literature in the right panel of Fig.~\ref{fig:CSkernel}: phenomenological MAP22 \cite{Bacchetta:2022awv}, ART23 \cite{Moos:2023yfa},  SV19 \cite{Scimemi:2019cmh} and lattice LPC22 \cite{LatticePartonLPC:2022eev,Deng:2022gzi},  SVZES \cite{Schlemmer:2021aij},  ETMC21 \cite{Li:2021wvl} and qTMDWF \cite{Bollweg:2025iol}. 
In contrary to other  approaches, the PB extractions do not assume any parametrization: different slopes of the extracted curves are consequences of the different treatments of radiation. The spread of the PB results covers a significant fraction of other extractions. The result with a flattening behaviour at large $b$ is especially interesting since most of the extractions in the literature assume a rising behaviour.

\paragraph{Remark on consequences of neglecting the non-perturbative Sudakov}
With dynamical $z_M$, the non-perturbative Sudakov $\Delta_a^{\textrm{NP}}$ is neglected. It has interesting phenomenological consequences, one of them is illustrated in Fig.~\ref{fig:CSkernel}. Another striking example is the center-of-mass energy $\sqrt{s}$ (in)dependence of the intrinsic-kt. Standard MC generators find intrinsic-kt parameter to be strongly dependent on $\sqrt{s}$ \cite{CMS:2024goo} whereas with the PB approach such a dependence is not observed \cite{Bubanja:2023nrd}. The reason for that was tracked down to the treatment of soft gluons via $\alpha_s$ and the non-perturbative Sudakov \cite{Bubanja:2024puv,Hautmann:2025fkw}. 
This illustrates that the known interplay of intrinsic transverse momentum with soft gluon resolution scale at the level of TMDs \cite{SadeghiBarzani:2024lnj,Bubanja:2023nrd} manifests itself also in physical observables and has important consequences for our predictive power.

\paragraph{Acknowledgements}
The presented results were obtained with A. Bermudez Martinez, F. Hautmann, L. Keersmaekers, M. Mendizabal Morentin, S. Taheri Monfared, A.M. van Kampen. We thank H. Jung for many discussions. \\
A. Lelek acknowledges funding by Research Foundation-Flanders (FWO) (1272421N, 1278325N).

\end{document}